\input harvmac
\def\sssec#1{$\underline{\rm #1}$}

\def\IB{\relax\hbox{$\inbar\kern-.3em{\rm B}$}}
\def\IC{\relax\hbox{$\inbar\kern-.3em{\rm C}$}}
\def\ID{\relax\hbox{$\inbar\kern-.3em{\rm D}$}}
\def\IE{\relax\hbox{$\inbar\kern-.3em{\rm E}$}}
\def\IF{\relax\hbox{$\inbar\kern-.3em{\rm F}$}}
\def\IG{\relax\hbox{$\inbar\kern-.3em{\rm G}$}}
\def\IGa{\relax\hbox{${\rm I}\kern-.18em\Gamma$}}
\def\IH{\relax{\rm I\kern-.18em H}}
\def\IK{\relax{\rm I\kern-.18em K}}
\def\IL{\relax{\rm I\kern-.18em L}}
\def\IP{\relax{\rm I\kern-.18em P}}
\def\IR{\relax{\rm I\kern-.18em R}}
\def\IZ{\relax\ifmmode\mathchoice
{\hbox{\cmss Z\kern-.4em Z}}{\hbox{\cmss Z\kern-.4em Z}}
{\lower.9pt\hbox{\cmsss Z\kern-.4em Z}}
{\lower1.2pt\hbox{\cmsss Z\kern-.4em Z}}\else{\cmss Z\kern-.4em Z}\fi}
\def\II{\relax{\rm I\kern-.18em I}}

\def\CA{{\cal A}}
\def\CB {{\cal B}}

\def\CD {{\cal D}}
\def\CE {{\cal E}}
\def\CF {{\cal F}}
\def\CG {{\cal G}}
\def\CH {{\cal H}}

\def\CK {{\cal K}}
\def\CL {{\cal L}}
\def\CM {{\cal M}}
\def\CN {{\cal N}}
\def\CO {{\cal O}}

\def\CS {{\cal S}}

\def\CV {{\cal V}}

\def\CX {{\cal X}}


\def\p{\partial}
\def\pb{\bar{\partial}}
\def\ib{{\bar i}}
\def\jb{{\bar j}}

\def\zb {\bar{z}}


\def\Tr{{\rm Tr}}

\def\sdtimes{\mathbin{\hbox{\hskip2pt\vrule height 4.1pt depth -.3pt
width
.25pt
\hskip-2pt$\times$}}}
\def\p{\partial}
\def\pb{\bar{\partial}}

\def\lieg{{\underline{\bf g}}}

\def\lies{{\underline{\bf s}}}
\def\inbar{\,\vrule height1.5ex width.4pt depth0pt}
\font\cmss=cmss10 \font\cmsss=cmss10 at 7pt
\def\sdtimes{\mathbin{\hbox{\hskip2pt\vrule
height 4.1pt depth -.3pt width .25pt\hskip-2pt$\times$}}}



\def\unlockat{\catcode`\@=11}
\def\lockat{\catcode`\@=12}

\unlockat

\def\newsec#1{\global\advance\secno by1\message{(\the\secno. #1)}
\global\subsecno=0\global\subsubsecno=0\eqnres@t\noindent
{\bf\the\secno. #1}
\writetoca{{\secsym} {#1}}\par\nobreak\medskip\nobreak}
\global\newcount\subsecno \global\subsecno=0
\def\subsec#1{\global\advance\subsecno
by1\message{(\secsym\the\subsecno. #1)}
\ifnum\lastpenalty>9000\else\bigbreak\fi\global\subsubsecno=0
\noindent{\it\secsym\the\subsecno. #1}
\writetoca{\string\quad {\secsym\the\subsecno.} {#1}}
\par\nobreak\medskip\nobreak}
\global\newcount\subsubsecno \global\subsubsecno=0
\def\subsubsec#1{\global\advance\subsubsecno by1
\message{(\secsym\the\subsecno.\the\subsubsecno. #1)}
\ifnum\lastpenalty>9000\else\bigbreak\fi
\noindent\quad{\secsym\the\subsecno.\the\subsubsecno.}{#1}
\writetoca{\string\qquad{\secsym\the\subsecno.\the\subsubsecno.}{#1}}
\par\nobreak\medskip\nobreak}

\def\subsubseclab#1{\DefWarn#1\xdef
#1{\noexpand\hyperref{}{subsubsection}%
{\secsym\the\subsecno.\the\subsubsecno}%
{\secsym\the\subsecno.\the\subsubsecno}}%
\writedef{#1\leftbracket#1}\wrlabeL{#1=#1}}
\lockat

\font\manual=manfnt \def\dbend{\lower3.5pt\hbox{\manual\char127}}


\def\boxit#1{\vbox{\hrule\hbox{\vrule\kern8pt
\vbox{\hbox{\kern8pt}\hbox{\vbox{#1}}\hbox{\kern8pt}}
\kern8pt\vrule}\hrule}}
\def\mathboxit#1{\vbox{\hrule\hbox{\vrule\kern8pt\vbox{\kern8pt
\hbox{$\displaystyle #1$}\kern8pt}\kern8pt\vrule}\hrule}}


%
\lref\pauldan{L.~Alvarez-Gaume, Commun.Math.Phys.{\bf 90} (1983)161\semi
D.~Friedan, P.~Windey, Nucl.Phys. B235 (1984) 395 }
\lref\ashvarz{A.S.~Schwarz, Lett. Math. Phys. {\bf 2} (1978) 247}
\lref\blzh{A. Belavin, V. Zakharov, ``Yang-Mills Equations as inverse
scattering
problem''Phys. Lett. B73, (1978) 53}
\lref\nikitathes{N.~Nekrasov, PhD. Thesis, Princeton 1996\semi
``Five-dimensional gauge theories and
relativistic integrable systems'', hep-th/9609219}


\lref\AHS{M.~ Atiyah, N.~ Hitchin and I.~ Singer, ``Self-Duality in
Four-Dimensional
Riemannian Geometry", Proc. Royal Soc. (London) {\bf A362} (1978)
425-461.}
\lref\fmlies{M. F. Atiyah and I. M. Singer,
``The index of elliptic operators IV,'' Ann. Math. {\bf 93}(1968)119}
\lref\bagger{E. Witten and J. Bagger, Phys. Lett.
{\bf 115B}(1982)202}
\lref\aniemi{A.~Hietamaki, A.~Niemi, A.~Morozov, ``Geometry Of $\CN=1/2$ Supersymmetry And The Atiyah-Singer
Theorem'', Phys.Lett. {\bf B263} (1991) 417 }
\lref\bjsv{ M. Bershadsky, A. Johansen, V. Sadov and C. Vafa,
``Topological Reduction of 4D SYM to 2D $\sigma$-Models'',
hep-th/9501096,
Nucl. Phys. {\bf B}448 (1995) 166}
\lref\BlThlgt{M.~ Blau and G.~ Thompson, ``Lectures on 2d Gauge
Theories: Topological Aspects and Path
Integral Techniques", Presented at the
Summer School in Hogh Energy Physics and
Cosmology, Trieste, Italy, 14 Jun - 30 Jul
1993, hep-th/9310144.}
\lref\bpz{A.A. Belavin, A.M. Polyakov, A.B. Zamolodchikov,
``Infinite conformal symmetry in two-dimensional quantum
field theory,'' Nucl.Phys.B241:333,1984}
\lref\braam{P.J. Braam, A. Maciocia, and A. Todorov,
``Instanton   moduli as a novel map from tori to
K3-surfaces,'' Inven. Math. {\bf 108} (1992) 419}
\lref\cllnhrvy{C.~Callan and J.~Harvey, Nucl. Phys. {\bf B250}(1985)427}
\lref\CMR{S. Cordes, G. Moore, and S. Ramgoolam,
`` Lectures on 2D Yang Mills theory, Equivariant
Cohomology, and Topological String Theory,''
hep-th/9411210, or see http://xxx.lanl.gov/lh94,
Nucl. Phys. Proc. Suppl. 41 (1995)}
\lref\devchand{Ch. Devchand and V. Ogievetsky,
``Four dimensional integrable theories,'' hep-th/9410147}
\lref\devchandi{
Ch. Devchand and A.N. Leznov,
``B \"acklund transformation for supersymmetric self-dual theories
for
semisimple
gauge groups and a hierarchy of $A_1$ solutions,'' hep-th/9301098,
Commun. Math. Phys. {\bf 160} (1994) 551}
\lref\dnld{S. Donaldson, ``Anti self-dual Yang-Mills
connections over complex  algebraic surfaces and stable
vector bundles,'' Proc. Lond. Math. Soc,
{\bf 50} (1985)1}
\lref\DoKro{S.K.~ Donaldson and P.B.~ Kronheimer,
{\it The Geometry of Four-Manifolds},
Clarendon Press, Oxford, 1990.}
\lref\donii{S. Donaldson, Duke Math. J. , {\bf 54} (1987) 231. }
\lref\fs{L. Faddeev and S. Shatashvili, Theor. Math. Fiz., 60 (1984)
206}
\lref\fsi{ L. Faddeev, Phys. Lett. B145 (1984) 81.}
\lref\fz{I. Frenkel, I. Singer, unpublished.}
\lref\fk{I. Frenkel and B. Khesin, ``Four dimensional
realization of two dimensional current groups,'' Yale
preprint, July 1995, to appear in Commun. Math. Phys.}
\lref\gmps{A. Gerasimov, A. Morozov, M. Olshanetskii,
 A. Marshakov, S. Shatashvili ,``
Wess-Zumino-Witten model as a theory of
free fields,'' Int. J. Mod. Phys. A5 (1990) 2495-2589
 }
\lref\vrldy{R.~ Dijkgraaf, E.~ Verlinde and H.~ Verlinde,
``Counting Dyons in $N=4$ String Theory'', CERN-TH/96-170,
hepth/9607}
\lref\vrlsq{E. Verlinde and H. Verlinde,
``Conformal Field Theory and Geometric Quantization,''
in {\it Strings '89},Proceedings
of the Trieste Spring School on Superstrings,
3-14 April 1989, M. Green, et. al. Eds. World
Scientific, 1990}
\lref\vrldglo{E. Verlinde, ``Global Aspects of
Electric-Magnetic Duality,'' hep-th/9506011}
\lref\ver{E. Verlinde, Nucl. Phys. {\bf B} 300 (1988) 360}

\lref\gerasimov{A.~ Gerasimov, ``Localization in GWZW and Verlinde
formula'', UUITP 16/1993, hepth/9305090}
\lref\jaap{J. Kalkman, Commun.Math.Phys. 153 (1993) 447}
\lref\BlThlgt{M.~ Blau and G.~ Thompson, ``Lectures on 2d Gauge
Theories: Topological Aspects and Path
Integral Techniques", Presented at the
Summer School in Hogh Energy Physics and
Cosmology, Trieste, Italy, 14 Jun - 30 Jul
1993, hep-th/9310144.}
\lref\btverlinde{M.~ Blau, G.~ Thomson,
``Derivation of the Verlinde Formula from Chern-Simons Theory and the
$G/G$
   model'',Nucl. Phys. {\bf B}408 (1993) 345-390 }
\lref\btqmech{M.~ Blau , G.~ Thompson, ``Topological gauge theories
from supersymmetric quantum mechanics on
spaces of connections'', Int. J. Mod. Phys. A8 (1993) 573-586}
\lref\wittgr{E.~ Witten, ``The Verlinde Algebra And The Cohomology Of
The Grassmannian'',  hep-th/9312104}

\lref\gottsh{L. Gottsche, Math. Ann. 286 (1990)193}
\lref\gothuy{L. G\"ottsche and D. Huybrechts,
``Hodge numbers of moduli spaces of stable
bundles on $K3$ surfaces,'' alg-geom/9408001}
\lref\GrHa{P.~ Griffiths and J.~ Harris, {\it Principles of
Algebraic
geometry},
p. 445, J.Wiley and Sons, 1978. }
\lref\ripoff{I. Grojnowski, ``Instantons and
affine algebras I: the Hilbert scheme and
vertex operators,'' alg-geom/9506020.}

\lref\hitchin{N. Hitchin, ``Polygons and gravitons,''
Math. Proc. Camb. Phil. Soc, (1979){\bf 85} 465}

\lref\hklr{N.~Hitchin, Karlhede, Lindstrom, and M.~Rocek,
``Hyperkahler metrics and supersymmetry,''
Commun. Math. Phys. {\bf 108}(1987)535}
\lref\hirz{F. Hirzebruch and T. Hofer, Math. Ann. 286 (1990)255}
\lref\hms{hep-th/9501022,
 Reducing $S$- duality to $T$- duality, J. A. Harvey, G. Moore and A.
Strominger}
\lref\johansen{A. Johansen, ``Infinite Conformal
Algebras in Supersymmetric Theories on
Four Manifolds,'' hep-th/9407109}
\lref\kronheimer{P. Kronheimer, ``The construction of ALE spaces as
hyper-kahler quotients,'' J. Diff. Geom. {\bf 28}1989)665}
\lref\kricm{P. Kronheimer, ``Embedded surfaces in
4-manifolds,'' Proc. Int. Cong. of
Math. (Kyoto 1990) ed. I. Satake, Tokyo, 1991}
\lref\amsi{B.S.~Acharya, M.~O'Loughlin, B.~Spence, ``Higher dimensional
analogues of Donaldson-Witten theory'', hep-th/9705138}
\lref\afsl{B.S.~Acharya, J.M.~Figueroa-O'Farill, B.~Spence, M.O'Loughlin,
hep-th/9707118}
\lref\KN{Kronheimer and Nakajima,  ``Yang-Mills instantons
on ALE gravitational instantons,''  Math. Ann.
{\bf 288}(1990)263}
\lref\krmw{P. Kronheimer and T. Mrowka,
``Gauge theories for embedded surfaces I,''
Topology {\bf 32} (1993) 773,
``Gauge theories for embedded surfaces II,''
preprint.}
\lref\dub{M.~Dubois-Violette, Journal of Geometry and Physics {\bf 19}
(1996) 18-30}
\lref\avatar{A. Losev, G. Moore, N. Nekrasov, S. Shatashvili,
``Four-Dimensional Avatars of 2D RCFT,''
hep-th/9509151.}
\lref\cocycle{A. Losev, G. Moore, N. Nekrasov, S. Shatashvili,
``Central Extensions of Gauge Groups Revisited,''
hep-th/9511185.}
\lref\clash{A.~Losev, G.~Moore, N.~Nekrasov, S.~Shatashvili
       ``Chiral Lagrangians, Anomalies, Supersymmetry, and Holomorphy'',
hep-th/9606082, Nucl.Phys. B484 (1997) 196-222}
\lref\maciocia{A. Maciocia, ``Metrics on the moduli
spaces of instantons over Euclidean 4-Space,''
Commun. Math. Phys. {\bf 135}(1991) , 467}
\lref\mickold{J. Mickelsson, CMP, 97 (1985) 361.}
\lref\mick{J. Mickelsson, ``Kac-Moody groups,
topology of the Dirac determinant bundle and
fermionization,'' Commun. Math. Phys., {\bf 110} (1987) 173.}

\lref\milnor{J. Milnor, ``A unique decomposition
theorem for 3-manifolds,'' Amer. Jour. Math, (1961) 1}
\lref\taming{G. Moore and N. Seiberg,
``Taming the conformal zoo,'' Phys. Lett.
{\bf 220 B} (1989) 422}
\lref\whatwethink{V.~Fock, N.~Nekrasov, A.~Rosly, K.~Selivanov,
``What We Think
About The Higher Dimensional Chern-Simons Theories'', Proc. of
Sakharov Conference, 1991}
\lref\nair{V.P.Nair, ``K\"ahler-Chern-Simons Theory'', hep-th/9110042
}
\lref\ns{V.P. Nair and Jeremy Schiff,
``Kahler Chern Simons theory and symmetries of
antiselfdual equations'' Nucl.Phys.B371:329-352,1992;
``A Kahler Chern-Simons theory and quantization of the
moduli of antiselfdual instantons,''
Phys.Lett.B246:423-429,1990,
``Topological gauge theory and twistors,''
Phys.Lett.B233:343,1989}
\lref\nakajima{H. Nakajima, ``Homology of moduli
spaces of instantons on ALE Spaces. I'' J. Diff. Geom.
{\bf 40}(1990) 105; ``Instantons on ALE spaces,
quiver varieties, and Kac-Moody algebras,'' preprint,
``Gauge theory on resolutions of simple singularities
and affine Lie algebras,'' preprint.}
\lref\nakheis{H.Nakajima, ``Heisenberg algebra and Hilbert schemes of
points on
projective surfaces ,'' alg-geom/9507012}
\lref\ogvf{H. Ooguri and C. Vafa, ``Self-Duality
and $N=2$ String Magic,'' Mod.Phys.Lett. {\bf A5} (1990) 1389-1398;
``Geometry
of$N=2$ Strings,'' Nucl.Phys. {\bf B361}  (1991) 469-518.}
\lref\park{J.-S. Park, ``Holomorphic Yang-Mills theory on compact
Kahler
manifolds,'' hep-th/9305095; Nucl. Phys. {\bf B423} (1994) 559;
J.-S.~ Park, ``$N=2$ Topological Yang-Mills Theory on Compact
K\"ahler
Surfaces", Commun. Math, Phys. {\bf 163} (1994) 113;
S. Hyun and J.-S.~ Park, ``$N=2$ Topological Yang-Mills Theories and Donaldson
Polynomials", hep-th/9404009}
\lref\parki{S. Hyun and J.-S. Park,
``Holomorphic Yang-Mills Theory and Variation
of the Donaldson Invariants,'' hep-th/9503036}
\lref\pohl{Pohlmeyer, Commun.
Math. Phys. {\bf 72}(1980)37}
\lref\pwf{A.M. Polyakov and P.B. Wiegmann,
Phys. Lett. {\bf B131}(1983)121}

\lref\wittjones{E. Witten, ``Quantum field theory and the Jones
polynomial,'' Commun.  Math. Phys., 121 (1989) 351. }
\lref\sen{A. Sen,
hep-th/9402032, Dyon-Monopole bound states, selfdual harmonic
forms on the multimonopole moduli space and $SL(2,Z)$
invariance,'' }
\lref\shatashi{S. Shatashvili,
Theor. and Math. Physics, 71, 1987, p. 366}
\lref\thooft{G. 't Hooft , ``A property of electric and
magnetic flux in nonabelian gauge theories,''
Nucl.Phys.B153:141,1979}
\lref\vafa{C. Vafa, ``Conformal theories and punctured
surfaces,'' Phys.Lett.199B:195,1987 }
\lref\VaWi{C.~ Vafa and E.~ Witten, ``A Strong Coupling Test of
$S$-Duality",
hep-th/9408074.}
\lref\vrlsq{E. Verlinde and H. Verlinde,
``Conformal Field Theory and Geometric Quantization,''
in {\it Strings'89},Proceedings
of the Trieste Spring School on Superstrings,
3-14 April 1989, M. Green, et. al. Eds. World
Scientific, 1990}
\lref\kvd{M. R.~Douglas, S.~Katz, C.~Vafa,
``Small Instantons, del Pezzo Surfaces and Type I' theory'', hep-th/9609071}
\lref\mwxllvrld{E. Verlinde, ``Global Aspects of
Electric-Magnetic Duality,'' hep-th/9506011}
\lref\wrdhd{R. Ward, Nucl. Phys. {\bf B236}(1984)381}
\lref\ward{Ward and Wells, {\it Twistor Geometry and
Field Theory}, CUP }
\lref\wittenwzw{E. Witten, ``Nonabelian bosonization in
two dimensions,'' Commun. Math. Phys. {\bf 92} (1984)455 }
\lref\grssmm{E. Witten, ``Quantum field theory,
grassmannians and algebraic curves,'' Commun.Math.Phys.113:529,1988}
\lref\wittjones{E. Witten, ``Quantum field theory and the Jones
polynomial,'' Commun.  Math. Phys., 121 (1989) 351. }
\lref\wittentft{E.~ Witten, ``Topological Quantum Field Theory",
Commun. Math. Phys. {\bf 117} (1988) 353.}
\lref\Witdgt{ E.~ Witten, ``On Quantum gauge theories in two
dimensions,''
Commun. Math. Phys. {\bf  141}  (1991) 153.}
\lref\Witfeb{E.~ Witten, ``Supersymmetric Yang-Mills Theory On A
Four-Manifold,'' J. Math. Phys. {\bf 35} (1994) 5101.}
\lref\Witr{E.~ Witten, ``Introduction to Cohomological Field
Theories",
Lectures at Workshop on Topological Methods in Physics, Trieste,
Italy,
Jun 11-25, 1990, Int. J. Mod. Phys. {\bf A6} (1991) 2775.}
\lref\wittabl{E. Witten,  ``On S-Duality in Abelian Gauge Theory,''
hep-th/9505186}
\lref\faddeevlmp{L. D. Faddeev, ``Some Comments on Many Dimensional Solitons'',
Lett. Math. Phys., 1 (1976) 289-293.}

\lref\douglas{M.R. Douglas, ``Enhanced Gauge
Symmetry in M(atrix) Theory,'' hep-th/9612126}
\lref\hs{J.A. Harvey and A. Strominger,
``The heterotic string is a soliton,''
hep-th/9504047
}
\lref\fa{G.~ Faltings, ``A proof of Verlinde formula'',
J.Alg.Geom.{\bf 3}, (1994) }
\lref\beil{A. Beilinson, V. Drinfeld,
``Quantization of
Hitchin's Integrable System and Hecke Eigensheaves'',
{\rm A. Beilinson's lectures at IAS, fall 1994}\semi
``Quantization of Hitchin's fibration and Langlands' program'', in
  `Algebraic and geometric methods in mathematical physics'
(Kaciveli, 1993), 3--7,
  Math. Phys. Stud., 19, Kluwer Acad. Publ., Dordrecht, 1996.}
\lref\sen{A. Sen, `` String- String Duality Conjecture In Six Dimensions And
Charged Solitonic Strings'',  hep-th/9504027}
\lref\oxbury{W.M. Oxbury, ``Spin Verlinde spaces and Prym theta functions'',
alg-geom/9602008 \semi
``Prym Varieties and the Moduli of Spin Bundles'',  alg-geom/941101}
\lref\thaddeus{M. Thaddeus,
``Stable pairs, linear systems and the Verlinde formula'',
alg-geom/9210007}
\lref\beauville{A. Beauville,
``Conformal blocks, fusion rules and the Verlinde formula'', alg-geom/9405001
\semi
``Vector bundles on curves and generalized theta functions:
recent results and open problems '', alg-geom/9404001}
\lref\szenesi{A. Szenes, ``The combinatorics of the Verlinde formulas'',
alg-geom/9402003}
\lref\szenesii{A.~Szenes, ``Iterated residues and Multiple Bernoulli
polynomials'', hep-th/9707114}
\lref\bks{L.~Baulieu, H.~Kanno, I.~Singer, ``Cohomological Yang-Mills
Theory in Eight Dimensions'', hep-th/9705127\semi
``Special Quantum Field Theories in Eight and Other Dimensions'',
hep-th/9704167}
\lref\cdfn{E.~Corrigan,
C.~Devchand, D.B.~Fairlie and J.~Nuyts, Nucl. Phys.{\bf B}214 (1983) 452}
\lref\Wittop{E.~Witten, ``Topological Quantum Field Theory'',
Comm. Math. Phys. {\bf 117}  (1988) 353}
\lref\BS{L.~Baulieu, I.~Singer, ``Topological Yang-Mills Symmetry'',
Nucl. Phys. {\bf B} (Proc. Suppl.) 5B (1988) 12-19}

\lref\rf{D.~Amati, S.~Elitzur, E.~Rabinovici, `` On induced gravity in
 two-dimensional
topological theories'', hep-th/9312003, Nucl.Phys.{\bf B}418 (1994) 45-80} 
\lref\elitzur{S.~Elitzur, G.~Moore,
A.~Schwimmer, and N.~Seiberg,
``Remarks on the Canonical Quantization of the Chern-Simons-
Witten Theory,'' Nucl. Phys. {\bf B326}(1989)108 \semi
G. Moore and N. Seiberg,
``Lectures on Rational Conformal Field Theory'',
in {\it Strings'89}, Proceedings
of the Trieste Spring School on Superstrings,
3-14 April 1989, M. Green, et. al. Eds. World
Scientific, 1990}
\lref\fadsha{L.~Faddeev, S.~Shatashvili, `` Algebraic  and  Hamiltonian 
Methods  in the theory of non-abelian anomalies'', 
Teor.Mat.Fiz. 60 (1984) 206-217,  english version
Theor.Math.Phys. 60 (1985) 770-778}

\lref\baugros{L.~Baulieu, B.~Grossman, ``A topological interpretation
of stochastic quantizations'', Phys.Lett. 212B (1988) 351\semi
L.~Baulieu, B.~Grossman, ``Monopoles and topological field theory'',
Phys.Lett. B214 (1988) 223}
\lref\bau{L.~Baulieu, ``Chern-Simons Three-Dimensional and Yang-Mills-Higgs Two Dimensional
Systems as Four Dimensional Topological Field Theories'', Phys. Lett. {\bf B232} (1989) 473}
\Title{ \vbox{\baselineskip12pt\hbox{hep-th/9707174}
\hbox{ITEP-TH-34/97}
\hbox{HUTP-97/A035}
\hbox{LPTHE-9733}}}
{\vbox{
\centerline{Chern-Simons And Twisted Supersymmetry}
\centerline{in Various Dimensions}}}
\medskip
\centerline{Laurent Baulieu\foot{URA 280 CNRS, associe aux Universit\'es
 Paris $VI-VII$}, Andrei Losev $^2$,
Nikita Nekrasov $^3$}

\vskip 0.5cm
\centerline{$^{1,2,3}$ LPTHE, Universit\'e Paris VI, Tour 16 - $1^{er}$
etage,4, Place Jussieu, 75252
Paris C\'edex 05}
\centerline{$^{2,3}$ Institute of Theoretical and Experimental
Physics,
117259, Moscow, Russia}
\centerline{$^{2}$ Department of Physics, Yale University,
New Haven, CT  06520, Box 208120}
\centerline{$^{3}$ Lyman Laboratory of Physics,
Harvard University, Cambridge, MA 02138}
\vskip 0.1cm
\centerline{baulieu@parthe.lpthe.jussieu.fr}
\centerline{nikita@string.harvard.edu }
\centerline{losev@waldzell.physics.yale.edu}

\medskip
\noindent
We introduce special supersymmetric
gauge theories in three, five, seven and nine
dimensions, whose compactification on two-, four-, six- and eight-folds
produces a supersymmetric quantum mechanics on moduli spaces
of holomorphic bundles and/or solutions to the
analogues of instanton equations in higher dimensions. The theories may
occur on the worldvolumes of $D$-branes wrapping manifolds of special
holonomy. We also discuss the theories with matter.

\Date{July  1997}

\newsec{Introduction}

Recent advances in string duality and Matrix theory in particular
suggest the existence of interesting theories in dimensions
higher then four, whose effective description at low energies
is that of a supersymmetric gauge theory.
The standard lore says that the gauge theory in the space-time of
dimension
higher then four is either (infrared) trivial
and/or non-renormalizable and therefore
does not exist as a field theory. One may study the gauge theories whose
ultraviolet description is provided by string theory. For example, the
physics of $D$-branes is described at low energies by the
supersymmetric gauge theory. This argument indicates that a restricted
set of correlation functions of gauge theory can be defined
even in the higher dimensional theories.

We attempt to describe three classes of
such theories in this
paper. The theories of the first type are the Cohomological Field Theories
(CohFT) \Wittop\BS\ describing intersection theory on
 a moduli space of solutions to some gauge covariant
equations $\Phi_{\alpha} = 0$
for a $D$-dimensional gauge field $A_{\mu}$ and possibly
scalar fields in adjoint representation. The space of gauge fields
and possible scalars is denoted as $\CA_{n}$, where $n$ denotes
topological sector. In gauge theory one usually sums over
all topological sectors. Let $\CA = \amalg_{n} \CA_{n}$. The set $\Phi$ of
equations  $\Phi_{\alpha}$ which can be
called ``topological gauge conditions'' define $\CG$-invariant
submanifold of $\CA$, where $\CG = \amalg_{n} \CG_{n}$ is the gauge group.
Suppose that the quotient $\CM_{n}$
of the space of solutions
of the system $\Phi = 0$ by the gauge transformations is finite dimensional
in each topological sector $n$.
The space $\CM_{n}$
depends on the choice of data entering $\Phi_{\alpha}$, such as the
space-time
manifold $\CX^{D}$, metric and/or any other geometrical object on $\CX^{D}$.
The theory is called $\CH_{D} ( {\CA}, \Phi, \CG)$.
Assume that gauge theory (in string theory context) defines a
compactification
of $\CM_{n}$. The correlation functions in the
theory $\CH_{D} ({\CA}, {\Phi}, {\CG})$
are the integrals of certain differential forms
$\omega_{i}$ over the space $\CM_{n}$:
\eqn\dcorr{
\langle \CO_{1} \ldots \CO_{p}\rangle = \sum_{n}
\int_{\CM_{n}} \omega_{1} \wedge
\ldots \wedge \omega_{p}
}
The definition of the operators $\CO_{i}$ and of the map
$\CO_{i} \mapsto \omega_{i}$. is provided by the field theoretic
realization of $\CH_{D}({\CA}, {\Phi}, {\CG})$.
Also, a  class of
Lagrangians is associated to $\CH_{D}({\CA}, \Phi,
\CG)$.

Suppose the theory $\CH_{D}({\CA}, {\Phi} , {\CG})$ is given. One may
define the theory called ${\CK}_{D}({\CA}, {\Phi}, {\CG}; R)$. It is $D+1$
dimensional field theory compactified on a circle of radius $R$.
\foot{Some of the ideas
presented here are explained in details in \nikitathes} Its input
is the same triple $({\CA}, {\Phi}, {\CG})$ as of the theory $\CH_{D}$.
The output is the set of correlation functions:
\eqn\dpocorr{
\langle \tilde \CO_{1} \ldots \tilde \CO_{p} \rangle =
\int_{\CM_{n}} {\hat A}_{R}({\CM}_{n}) \omega_{1} \wedge \ldots \wedge
\omega_{p}}
where
the radius of the circle
is the expansion parameter of the $\hat A$ genus:
\eqn\aroof{
{\hat A}_{R}({\CM}_{n}) = \prod_{i=1}^{{\half} {\rm dim}{\CM_{n}}} {{R
x_{i}/2}\over{{\rm sinh}(R x_{i}/2)}},}
where $x_{i}$'s are the Chern roots of the tangent bundle to
${\CM}_{n}$\foot{It may seem that knowledge of \dcorr\ allows one to
compute \dpocorr\ immediately. In fact, the subtleties with
compactification of $\CM_{n}$ make the problem unaccessible
to current techniques. Moreover it is not
clear whether $A$-genus is among the
observables of $D$-dimensional
theory.
That is why we call it a new theory.}
Superficially the construction of $\CK_{D}$ is similar to that of
$\CH_{D+1} ( L{\CA}, L{\Phi}, L{\CG})$ where $L\CA$, $L\CG$ denote
respectively the loop space of $\CA$ and the loop group of $\CG$.
The equations $L{\Phi}$ are the same equations $\Phi$, imposed at
each point of a loop separately. The theory $\CH_{D+1}(L{\CA}, L{\Phi},
L{\CG})$
is sick since its moduli space is the loop space of the moduli space
$\CM$. The theory $\CK_{D}$ is defined by enhancing the symmetry group
to
$\IG = L\CG \sdtimes U(1)$, where $U(1)$ acts by rotations of loops.
The difference with ordinary CohFT's is that this $U(1)$ is treated as a
global
symmetry. In particular, the ghost for ghost $k = {1\over{R}}$
associated to $U(1)$
is fixed rather then integrated over as it is done in the ordinary
case. This
definition can be illustrated by the
compactification of the theory on a $D$-fold which
reduces the model to supersymmetric quantum mechanics on the
moduli space $\CM_{n}$.

The last theory in our list is
${\CE}ll_{D}({\CA}, {\Phi}, {\CG}; {\rho} , {\tau})$
which is associated to the spaces of double loops, i.e. of the maps of torus
$E_{\rho, \tau}$ to $\CA$. It is $D+2$-dimensional
theory.
The correlation function  in ${\CE}ll_{D}$
are related to the elliptic genera of
$\CM$ \nikitathes.

We discuss the examples of $\CK_{D}$-theories for $D=8,6$
related to supersymmetric Yang-Mills theories in
seven and nine
dimensions.
They are twisted versions of dimensional reductions of the
$\CN=1$ $d=10$ superYang-Mills theories.
The $d=10$ theory  is the theory of type ${\CE}ll_{8}$.
We also describe briefly the theories $\CK_{D}$ in dimensions five
and three.  Superficially, $9d$ and $7d$
theories are related to the octonionic structure which prevails in $8$
dimensions, $5d$ is related to quaternions and $3d$ to complex numbers.

One  motivation for studying such theories is  the following. Recently
the higher dimensional analogues of Donaldson-Witten theory
were studied \bks. It is natural to ask whether the analogues of
Chern-Simons
theory (as ``theories of A.~Schwarz type'' \ashvarz)
exist. We will find them in the framework of $\CK_{D}$ theories.

If the theory is defined in $D+1$ dimensions one may try to
study it on $\CX^{D} \times I$,
rather then $\CX^{D} \times S^{1}$. We show that this
leads to an interesting $WZW$-like theory in $D$ dimensions.

Concluding the introduction we must warn the reader that we do not
discuss various subtleties related to our choice of regularization.
In particular, not every gauge group may be realized in nine and
eight dimensions
within string theory, according to today's knowledge. It is very
interesting
to see how
the subtleties of higher dimensional physics are
reflected in topology (anomalies) and geometry of
the moduli spaces $\CM$.

The paper is organized as follows. The chapter $2$ is devoted to the
theories $\CH_{D}$. The chapter $3$ reviews some important constructions
in supersymmetric quantum mechanics and then apply them in
infinite-dimensional context. This application yields the
theory $\CK_{D}$. Then we discuss the examples of
theories, related to octonions. The chapter $4$ is devoted
to observables. We find that Chern-Simons functionals
can be promoted to the {\it bona fide} obsrevables
in the theory $\CK_{D}$. The chapter $5$ briefly described the theories
in three and five dimensions, and also remarks on the theories
with matter. The chapter $6$ deals with $WZW$-like theories
in higher dimensions. We present our conclusions in the chapter $7$.

\newsec{Constructions of the moduli
spaces: $\CH_{D}$ Theories}

\lref\oxbury{W.M. Oxbury, ``Spin Verlinde spaces and Prym theta functions'',
alg-geom/9602008 \semi
``Prym Varieties and the Moduli of Spin Bundles'',  alg-geom/941101}
\lref\thaddeus{M. Thaddeus,
``Stable pairs, linear systems and the Verlinde formula'',
alg-geom/9210007}
\lref\beauville{A. Beauville,
``Conformal blocks, fusion rules and the Verlinde formula'', alg-geom/9405001
\semi
``Vector bundles on curves and generalized theta functions:
recent results and open problems '', alg-geom/9404001}
\lref\szenesi{A. Szenes, ``The combinatorics of the Verlinde formulas'',
alg-geom/9402003}
\lref\szenesii{A.~Szenes, ``Iterated residues and Multiple Bernoulli
polynomials'', hep-th/9707114}

\lref\cdfn{E.~Corrigan,
C.~Devchand, D.B.~Fairlie and J.~Nuyts, Nucl. Phys.{\bf B}214 (1983) 452}
\subsec{Cohomological Field Theories}

Cohomological Field Theories allow to construct a theory of
integration over a quotient ${\CM} = {\CN}/{\CG}$
of submanifold ${\CN}$ of a manifold $\CA$ by the action of a group $\CG$.
The physically interesting cases related to gauge theories
correspond to $\CA$ being a space of gauge fields in some
principal $G$-bundle $P$, $\CN$ being the set of zeroes of a section
$s$ of a certain infinite-dimensional vector bundle $\CV$ over $\CA$.

For example, in four dimensional gauge theory one may take:
$$
\CV = \Gamma \left( \Omega^{2,+}({\CX}^{4}) \otimes \lieg_{P} \right),
$$
where $\lieg_{P}$ is the associated to $P$ adjoint bundle. The natural section
$s$ is in this case $s = F_{A}^{+}$.

In general one sums over all topological types of $P$, hence
the quotient $\CM$ is the disjoint union of finite-dimensional
(this is an assumption) manifolds $\CM_{n}$:
$$
\CM = \amalg_{n} \CM_{n}
$$
The index $n$ stands for ${\rm Ch}(P)$ and $w_{*}(P)$.
Now assume that $\CG$-equivariant bundle $\CV$ over $\CA$ is given,
and choose a non-degenerate section $s$ of it.
More precisely,  the
linearization of the equation $s=0$ must be Fredholm on the complement
to the tangent space to the gauge orbit.
We think of the linearization of the equations as of the map:
$ds: T\CA \to \CV$ and it is required to have finite dimensional
kernel and cokernel on the complement to the tangent space to gauge orbit.
We endow  $\CV$  with a $\CG$-invariant
 metric $g_{\alpha\beta}$. Sometimes we write the section $s$ in components:
$s = \{ \Phi_{a} \}$. In the introduction we denoted by $\Phi$ the set
of equations $\Phi_{\alpha}$. It is more accurate to call $\Phi$ the
pair $(\CV, s)$ as it is this pair which enters the definition of
our theory
$\CH_{D}({\CA}, {\Phi}, {\CG})$.

We proceed
by introducing the standard package of
gauge CohFT \Witr.
One has classical gauge fields  $A_{\mu}, \ \ldots$,
topological gauge conditions $\Phi_{\alpha} = 0$
and gauge symmetries $A_{\mu} \mapsto
g^{-1} A_{\mu} g + g^{-1}\p_{\mu}g,  \ldots$, where $\ldots$ denote
 possible additional    fields and the action of the gauge group on them.

One has  also     fermions $\psi_{\mu}$
which  represent the exterior derivatives of $A_{\mu}$ and
the complex scalar field $\phi$ with values in the adjoint
representation which represents the degree two generator in equivariant
Cartan complex. In order to impose the equations $\Phi_{\alpha} = 0$
one needs a  multiplet of fields with opposite statistics taking values
in $\CV^{*}$
$(\chi^{\alpha}, H^{\alpha})$.
One also needs the projection multiplet $(\bar \phi, \eta)$
\Wittop\BS\CMR.
All these fields fit in the context of the BRST technology, and the main
process of building the QFT can be understood as a gauge fixing of the
symmetries
of relevant topological actions.

This leads one to introduce a  generator $Q$,  which is nilpotent
{\it up to the gauge transformations}:
\eqn\nilgen{\eqalign{& Q A_{\mu } = \psi_{\mu}, \quad  Q \psi_{\mu} =
D_{\mu} \phi \equiv \p_{\mu} \phi + [ A_{\mu}, \phi]\cr
& Q \phi=0\cr
& Q \chi^{\alpha} = H^{\alpha}, \quad Q H^{\alpha} = \phi^{a}
T^{\alpha}_{a, \beta} \chi^{\beta} \cr
& Q \bar \phi = \eta, \quad  Q \eta = [ \phi, \bar \phi] \cr}  }
where $T_{A,\beta}^{\alpha}$ represents the action of the gauge group in
the fibers of
 $\CV$. In all our examples the action will be simply
adjoint.

The property
\eqn\nil{Q^2= \quad {\rm   gauge\ transformation \ with\  parameter }
\  \phi }
allows to consider
  the  following $Q$-invariant and  gauge invariant Lagrangian:
\eqn\lagran{\CL = \int_{\CX} \{ Q, i \chi^{\alpha} (\Phi_{\alpha} + ie^{2}
g_{\alpha\beta} H^{\beta}) + {\Tr}
\left( \bar\phi  D^{\mu}  \psi_{\mu} + \eta [\phi, \bar\phi]\right) \} }
The gauge fixing interpretation of the  induced  QFT  is clear.
Formally, the path integral
$$
\int {{DA D\psi D\eta D\bar\phi D\phi}\over{\rm Vol(\CG)}} e^{-\CL} \ldots
$$
is $e^{2}$ independent and therefore reduces to the integral over $\CM =
\Phi^{-1}(0)/{\CG}$ provided that appropriate observables are inserted
in $\ldots$.
\lref\matqui{V.~Matthai and D.~Quillen, ``Superconnections, Thom classes
and Equivariant Differential Forms'', Topology {\bf 25} (1986) 85}
\lref\matblau{M.~Blau,
``The Mathai-Quillen Formalism and Topological Field Theory'',
hep-th/9203026 , J. Geom. Phys. 11 (1993) 95-127}
The integral is the infinite-dimensional version
of Matthai-Quillen \matqui\matblau\jaap\
representative of the Euler class of $\CV$.

Now let us discuss the choices of $\CV$ and $s$.
The curvature of the gauge field at a given point of space-time $\CX$
is an element of $\Lambda^{2} \otimes \lieg$ - a $D(D-1)/2 \times {\rm
dim}\lieg$ dimensional vector space. Let $V$ be some rank $D-1$
vector bundle over $\CX$ and choose a fiber-wise linear
map
$\Psi : \Omega^{2}({\CX}) \to V$.
We try as the bundle $\CV$ the space of sections of $V \otimes \lieg_{P}$:
$$
\CV = \Gamma (V \otimes \lieg_{P})
$$
Consider the equations
\eqn\asleq{s = \Psi \cdot F_{A} = 0.}
These equations lead to nice CohFT's,
provided that the complex:
\eqn\cmplx{\matrix{& & d_{A} & & \Psi \cdot d_{A} & & \cr
& \Omega^{0} \otimes \lieg_{P} & \to &
\Omega^{1} \otimes \lieg_{P} & \to &
V \otimes \lieg_{P}& \cr}}
is elliptic. Its index equals the virtual dimension of $\CM_{n}$.

Explicit examples of such equations are, e.g.,   flatness equations in $D=2$ : $F = 0$,  (anti-)self-duality in $D=4$: $F^{\pm} = 0$ and
 complexified instantons and octonionic instantons in $D=8$
and their dimensional reductions \cdfn, \bks, \amsi.
\lref\bsv{M. Bershadsky, V. Sadov, C. Vafa,
``D-Branes and Topological Field Theories'', hep-th/9511222,
Nucl.Phys. B463 (1996) 420-434}
Once the nice set of equations is obtained in dimension $D$ one may
get the equations in lower number of space-time dimensions
by performing dimensional reduction. In going to $D^{\prime}$ dimensions
the
space $\CA$ becomes a space of pairs $(A, H)$, where $A$ is the gauge field
in a bundle $P^{\prime}$ over $\CX^{D^{\prime}}$ and $H$ is the section of
$\lieg_{P^{\prime}} \otimes E$. The bundle $E$ has rank $D - D^{\prime}$. Its
topology may be rather involved.  Physics of Dirichlet-branes suggests that
$E$ may be interpreted as a normal bundle to $\CX^{D^{\prime}}$ in some
$D$-dimensional Calabi-Yau manifold \bsv.

In this way one gets Bogomolny equations in $D^{\prime} = 3$, Hitchin
equations in $D^{\prime}= 2$ and balanced version of Donaldson-Uhlenbeck-Yau
equations in $D^{\prime}=6$.

\newsec{  Theories in $D+1$ dimensions: $\CK_{D}$ theories}

In this chapter we generalize the ideas introduced
in \nikitathes, where the two and four dimensional
topological Yang-Mills theories
were related to three and five dimensional theories.
We shall see  the existence of supersymmetric  theories  in
$D+1$ and  $D+2$ dimensions
which can be projected to
the  $D$ dimensional topological theories discussed so far.
Eventually, we will see that they have other limits in $D$ dimensions,
of the WZW type.

\subsec{Supersymmetric quantum mechanics}

In this section we remind a few important constructions:
supersymmetric quantum mechanics with target $M$, the one with
target being a submanifold $N \subset M$ and the one with target being
a quotient $M/G$ by an action of a group $G$. These constructions
can be viewed as a passage from $D=0$ dimensional to $D=0+1$
dimensional theory.

 Supersymmetric $\CN = \half$ quantum mechanics is the way to
describe spinors on a manifold $M$ in the first quantized formalism.
One studies the integrals over the space of maps
$(x^{\mu}(t), \psi^{\mu}(t))$ of the worldline to the $(m \vert m)$
dimensional superspace $\Pi TM$. The path integral measure
$D x D\psi$ is well-defined and invariant under any changes
of the coordinates $x$, provided that they are accompanied
by the corresponding change in $\psi$.

The worldline  supersymmetry:
\eqn\smmtry{\eqalign{
\delta x^{\mu} = \psi^{\mu} \cr
\delta \psi^{\mu} = k \p_{t} x^{\mu} \cr}}
squares to the time translation with parameter  $k$:
$\delta^{2} = k \p_{t}$.
$\delta$ acts as nilpotent operator on the observables, invariant
under the rotation of the parameter $t$ and it is possible
to define  a cohomology space.
\lref\atiyah{M.~Atiyah, Index theorem and Duistermaat-Heckmann formula}
\lref\atbott{M.~Atiayh, R.~Bott, ``The moment map and equivariant
cohomology'',
Topology {\bf 23} (1984) 1}
The symmetry $\delta$ has the following
 interpretation.
Consider the space of parameterized loops $X = LM$. The differential forms
on $X$ can be identified with the functionals of $x^{\mu}(t)$
and $\psi^{\mu}(t)$, where $\psi^{\mu}(t)$ corresponds to the
differential $dx^{\mu}(t)$. The  group $U(1)$ acts on $X$ by rotations
of loops and $\delta$ is the equivariant derivative $d + k \iota_{V}$,
with $V$ representing the vector field
$\p_{t} x^{\mu} {{\delta}\over{\delta x^{\mu}}}$.
The number $k$
which serves as a normalization constant  is degree two generator
in $U(1)$ equivariant cohomologies \atbott.

The  universal
$\delta$-exact action which exists for any Riemannian $M$
is:
\eqn\rgltr{\eqalign{
\beta_{k} &=
\int dt g_{\mu\nu}\left( {\psi}^{\mu}  \nabla_{t}  \psi^{\nu} +
k \partial_{t} x^{\mu} \partial_{t} x^{\nu} \right)
\cr
&\sim
\delta \int dt \left( g_{\mu\nu} {\psi}^{\mu}  \p_{t} x^\nu\right)\cr
}}
where $\nabla_{t}$ is the pull-back of the Levi-Chivita connection on $TM$
to
the circle. The advantage
of having the fermions and symmetry $\delta$ is the possibility
to use the localization principle for evaluation of the
partition function. The fixed points
of the group action are the constant loops. Thus, the partition
function
\eqn\sqmprtn{
\int DxD\psi \quad e^{-\beta_{k}}
}
can be expressed as the integral over the space of
constant loops, i.e. $M$ itself. The integrand
is given by the ratio of determinants one gets by expanding around
the constant loop. It is well-known that the answer  is
the index of Dirac operator \pauldan\aniemi\foot{The
Dirac operator is the space-time interpretation of
$\delta$}.
The partition function is formally
independent of any $\delta$-exact
terms one could add to the action as long as everything
is invariant under the rotations of the circle
(and consequently $\delta$ squares to zero).

Suppose $\omega = {\half}\omega_{\mu\nu}dx^{\mu}\wedge dx^{\nu}$ is a closed
two-form on a simply-connected manifold $M$ with integer periods.
One can form
another   $\delta$-invariant
action:
\eqn\alp{
\alpha_{k} = \int_{S^{1}} dt
({\half} \omega_{\mu\nu} \psi^{\mu}
\psi^{\nu} + k  {\theta}_{\mu} \p_{t} x^{\mu})}
Here we introduced a one-form $\theta = \theta_{\mu}dx^{\mu}= d^{-1}
\omega$.
Of course, $\theta$ is only defined
locally, but the equivariant form $e^{2\pi ip \alpha}$ (it is sometimes called
Polyakov's loop) is
well-defined, provided $p k \in \IZ$.

When \alp\ and \rgltr\ are taken
together the answer for the partition function is the index of
Dirac operator coupled to abelian gauge field, whose curvature
is $2\pi i \omega$.

It is of interest of extending this formalism in two respects\foot{For the
extended supersymmetry the relevant construction was presented in \btqmech,
but here we need to treat the $\CN={\half}$ version of the story.}
 (we will need
both): the action of a group $G$ on $M$ and the quantization of a
submanifold $N$ of $M$. In the case of our interest the moduli space
$\CM$ is a quotient of a submanifold $N$ of a manifold $M$.
Suppose that $N$ can be realized as a set of zeroes of a section
$s = \{ \Phi_{\alpha} \}$ of some vector bundle $V$ over $M$.
To get a restriction onto submanifold $N$ one introduces a multiplet
of Lagrange multipliers $H^{\alpha}$ and their superpartners
$\chi^{\alpha}$
which are the sections of a pullback of the bundle $V^{*}$ to the loop.
Let $B_{\alpha}^{\beta} =B_{\mu,
\alpha}^{\beta} dx^{\mu}$ be a connection on $V$. We assume that $V$
is endowed with a metric $g^{\alpha\beta}$ and that $B$ is compatible
with it. Let $\CF = dB + B^{2}$ be the curvature of $B$.
The supersymmetry $\delta$ acts on $\chi, H$ is as follows:
\eqn\ghst{\eqalign{&\delta \chi^{\alpha}
= H^{\alpha} - B_{\mu, \alpha}^{\beta} \psi^{\mu} \chi_{\beta}\cr
& \delta H^{\alpha} = k \p_{t} \chi^{\alpha} -
B_{\mu,\beta}^{\alpha} ( H^{\beta}\psi^{\mu} - {\dot x}^{\mu} \chi^{\beta})
- {\half} \psi^{\mu}\psi^{\nu} \left(
{{\CF}_{\mu\nu}}\right)^{\alpha}_{\beta}\chi^{\beta}\cr}}
Consider the following interactions:
\eqn\inti{
\gamma_{k} = \delta \left( i\int_{S^{1}} \chi^{\alpha} \Phi_{\alpha}
\right) =
i \int H^{\alpha} \Phi_{\alpha} + i \int \chi^{\alpha} \psi^{\mu}
\nabla_{\mu} \Phi_{\alpha},}
where $\nabla_{\mu} \Phi_{\alpha}$ is:
$$
\nabla_{\mu} \Phi_{\alpha} = {{\p \Phi_{\alpha}}\over{\p x^{\mu}}} +
B_{\mu, \alpha}^{\beta} \Phi_{\beta},
$$
and
\eqn\intii{
\delta_{k} = \delta \left( \int_{S^{1}} - {\half} g_{\alpha\beta}
\chi^{\alpha}H^{\beta} \right) =  -
{\half} \int_{S^{1}}\left( H_{\beta}H^{\beta}  +
\chi_{\beta} \CD_{t} \chi^{\beta}  + \left( \CF_{\mu\nu}\right)_{\alpha\beta}
\psi^{\mu}\psi^{\nu}
\chi^{\beta}\chi^{\alpha} \right),  }
where the covariant derivative $\CD_{t}$ is
defined with the help of the pullback of $A$:
$$
\CD_{t} \chi^{\alpha} = k\p_{t} \chi^{\alpha} +
B_{\mu, \beta}^{\alpha} {\dot x}^{\mu} \chi^{\beta}.
$$
The path integral
\eqn\locpthint{\int Dx D\psi D\chi DH \exp ( \beta_{k} + \gamma_{k} +  e^{2}
\delta_{k} )}
reduces to the integral \sqmprtn\  for the submanifold $N$, as can be
seen by integrating out $H$  and then taking the limit $e^{2} \to 0$.
The reason why the prescription with $e^{2} \neq 0$ is better then
$e^{2}=0$ is that it works even if the section $s$ is not generic
(e.g. $s=0$).

If the group $G$ acts on $M$ then $LM$ is acted on by the
group $\IG = LG \sdtimes U(1)$, where $U(1)$ acts on $LG$ by rotations
of the loops. To avoid possible confusions let us stress that it is only
$LG$ which is being gauged, not $\IG$.
The appropriate setting is the equivariant
cohomology of $LM$ with respect to $\IG$.
The action of the group is incorporated by making $\delta$
the equivariant derivative with respect to $\IG$. One introduces
a field $\phi^{a}(t)$, which takes values in the Lie algebra $\lieg$ of $G$.
The field $\phi^{a}$ may be thought of  a one-dimensional gauge field.
Then new operator $\delta$  is the sum of \smmtry\ and the operator which
maps $\psi^{\mu}$ to $\phi^{a}(t) V_{a}^{\mu}(x(t))$
and $H^{\alpha}$ to $\phi^{a}(t) T_{a,\beta}^{\alpha} \chi^{\beta}$,
where $V^{\mu}_{a}$ is a vector field on $M$, representing
the Lie algebra element $T_{a}$, and $T_{a, \beta}^{\alpha}$ represents
the action of $G$ on the normal bundle to $N$. (Of course,
for the whole construction to work the submanifold $N$ must
be $G$-invariant).
The regulator \rgltr\ is gauged:
\eqn\nrgltr{\eqalign{
\beta_{k} & = \delta (\  \int dt g_{\mu\nu} \psi^{\mu} D_{t}
x^{\nu} )\cr
& = \int dt \left(  g_{\mu\nu} \psi^{\mu} ({\nabla}_{t} \psi^{\nu}
+ \phi^{a} \p_{\xi}V_{a}^{\nu}  \psi^{\xi} ) + k g_{\mu\nu}
D_{t}x^{\mu}D_{t}x^{\nu} \right)\cr
& \quad D_{t}x^{\mu} = \p_{t}x^{\mu} + \phi^{a}V_{a}^{\mu} \cr}}

The form \alp\ changes as follows. If $G$ preserves
$\omega$  and $\mu_{a}$ is the moment map, then
\alp\ is replaced by:
\eqn\nalp{
\alpha_{k} = \int_{S^{1}} dt
({\half} \omega_{\mu\nu} \psi^{\mu}
\psi^{\nu} + k  {\theta}_{\mu} \partial_{t} x^{\mu}
+ {\phi}^{a} \mu_{a})}

\subsec{The theory $\CK_{D}$: fields and supercharge}

We want to generalize these constructions
to cover infinite-dimensional cases. More precisely,
we wish to consider as $M$  the
space $\CA$
of gauge fields (not specifying the instanton sector)
on space-time manifold $\CX$, the group $G$ is the gauge
group and the submanifold $N$ is the space of solutions
of some natural (in particular, local in space-time)
equations $\Phi_{\alpha} (F_{A}) = 0$,
where  $F_{A}$ is the curvature of the gauge field.
Starting with a theory $\CH_{D}$  in $D$ dimensions,
defined   by equations    \nilgen\
and \lagran, we  define  a
related  interesting supersymmetric theory in $D+1$ dimensions
by means of the following procedure,
motivated
by the discussion of supersymmetric quantum mechanics.
The generator $Q$ below is simply the operator $\delta$ generalized
to the infinite-dimensional setting. For simplicty we set $k=1$.
It can be recovered by the rescaling of the radius of the $t$ circle.
We also
set the connection $B$ to zero.

 One considers the same fields as the previous section with   a dependence
on an additional  coordinate $t=x^{D+1}$, e.g., $A(x)\to A(x,t)$.
Moreover, one introduces an extra component $A_t (x,t)$ in
$A_\mu (x,t)$, and an
 anticommuting component  $\psi_t(x,t)$ in $\psi _\mu (x,t)$.
The fermion $\psi_{t}$ is playing the r\^ole of $\eta$ (see below).
Instead of complex field $\phi$ we have a real $\lieg$-valued scalar
$\varphi$.

  The operator $Q$ \nilgen\ is   generalized into:
\eqn\newsusy{\eqalign{& Q A_{\mu } = \psi_{\mu}, \quad  Q \psi_{\mu} =-
F_{\mu t}- i D_{\mu} \varphi \cr
& Q \chi^{\alpha} = H^{\alpha}, \quad Q H^{\alpha} = D_{t} \chi^{\alpha} +
i[\varphi, \chi^{\alpha} ] \cr
& Q \varphi  = i \psi_{t}, \quad Q \psi_{t} = -i D_{t} \varphi \cr}}
(the index  $\mu$ now runs from $1$ to $D+1$.)

The important property of $Q$ is:
\eqn\nil{Q^2=\p_{t} + {\rm   gauge\ transformation
\ with\ parameter }\ A_{t} + i\varphi }
One can   improve    \newsusy\  by introducing a scalar  ghost
$c$ (interpreted as an ordinary Faddeev-Popov ghost),
which gives rise to a modified  operator $Q$
squaring to $\p_{t}$ only. This is equivalent to doing the standard
Weil complex procedure \dub, \CMR.

We modify the transformation laws: $Q \to \lies$:
\eqn\susybrs{\eqalign{&\lies A_{\mu} = \psi_{\mu} + D_{\mu} c,\quad
{\lies}\varphi = -[c,\varphi ] +i \psi_{t}\cr
&\lies \psi_{\mu} = F_{t \mu  } - i D_{\mu}\varphi-[c, \psi_{\mu}], \quad
{\lies} \chi_{\alpha} = H_{\alpha} -[c, \chi_{\alpha}]\cr
&  \lies H_{\alpha}
= D_{t} \chi_{\alpha} + i [\varphi, \chi_{\alpha}]   -[c,   H_{\alpha}  ]\cr
&
 \lies c = A_{t} + i\varphi - {1\over{2}}[c,c]
\cr}}
(Notice that  $Q(A_t+i\varphi)=0$  and
$\lies (A_t+i\varphi)= \p_{t} c +[A_t+i\varphi, c]$).
As a result:
\eqn\nil{\lies ^2={\p_{t}}, }
and one can  properly deal   with the gauge fixing of the ordinary
gauge symmetry  of the
$Q-$invariant action, by adding a ${\lies}$-exact term.\foot{For example,
one may use Landau type gauges.}

The  equations \newsusy\susybrs\
break  $SO(D+1)$ invariance, as it is explicit in the
transformation law of $\psi_\mu$.
We will come back to this point when discussing observables.

It is useful  to establish the dictionary according to which the
transformation $\lies$ in \susybrs\
encodes the usual nilpotent topological BRST  operator of the $D$-
dimensional Yang-Mills theory.

For this, one   can decompose    the equations \susybrs\
  as ($1\leq i\leq D$):
\eqn\susybrs{\eqalign{
&\lies A_{i} = \psi_{i} + D_{i} c
\cr
&\lies c = A_{t} + i\varphi - {1\over{2}}[c,c]
\cr
 &\lies \psi_{i} = \p_{t} A_i
-D_{i}(A_t+i\varphi)
-[c, \psi_{i}]
\cr
&
\lies (A_t+i\varphi)= \p_{t} c -[c, A_t+i\varphi]
\cr
&
\lies (A_t-i\varphi)= 2\psi_{t}+\p_{t} c
-[c, A_t-i\varphi]
\cr
 &
 {\lies}(2\psi_{t}+\p_{t} c ) =
-[c,2\psi_{t}+\p_{t} c]
+[A_t+i\varphi, A_t-i\varphi]\cr
&
  {\lies}\chi_{\alpha} = H_{\alpha} -[c, \chi_{\alpha}]\cr
&   \lies H_{\alpha} =
D_{t} \chi_{\alpha} + i [\varphi, \chi_{\alpha}]  -[c,   H_{\alpha}  ]  } }

The relation between  the symmetries  in $D$ and $D+1$ dimensions is
that  $\phi$ and  $\bar \phi$  are  replaced
respectively  by $\p_{t} + A_{t} + i\varphi$  and
$\p_{t} + A_{t} - i\varphi$.
Moreover,
by performing the standard dimensional reduction in
 which all fields
become $t$ independent and
all terms involving $\p_{t} $ drop out one immediately  sees that
  $\phi=A_t+i\varphi$ can be interpreted in $D$ dimension as
the ghost of ghost for $c$, with ghost number $2$,  and
$\bar \phi=A_t-i\varphi$ is the antighost for antighost with ghost number
$-2$,
while $\eta = 2\psi_{t}+\p_{t} c$ is the
Lagrange multiplier with ghost number $-1$
for the gauge condition on the  vector ghost fields $\psi_i$.
Hence as compared to  $D$ dimensional theory
there is a violation of ghost number  in the $D+1$ dimensional theory.
Obviously, in the process of dimensional reduction, one gets
$\lies ^2=0$.

The Lagrangian of the $D+1$ dimensional theory is a straightforward
generalization of \lagran:
\eqn\newlagran{\CL = \int_{\CX^{D} \times \IR^{1}} \{ Q, i \chi^{\alpha}
(\Phi_{\alpha} + ie^{2}
g^{\alpha\beta} H_{\beta}) +
{\Tr} \psi^{\mu} (F_{t\mu} + i D_{\mu} \varphi) \} }

\subsec{The theory in $D+2$ dimensions: $\CE ll_{D}$.}

  In turn, on can go one more dimension higher, that is to $D+2$,
where the field $\varphi$ becomes the component
$A_{D+2}$ of the  $D+2$ dimensional gauge field $A$.
Straightforward computations indicate
that one  obtains a theory with a supersymmetry charge satisfying\foot{As
it has been noticed in
the discussions with G.~Moore and S.~Shatashvili there are difficulties
with defining Chern-Simons like observables in this $D+2$ dimensional
theory}:
\eqn\nil{\lies ^2={{\p}_{D+1}}
+i{\p}_{D+2} = \p_{\bar z}, \quad z = {\half} (x^{D+1} + i x^{D+2})}

The theory in $D+2$ dimensions is likely to be untwisted to an  ordinary
$\CN=1$ supersymmetric theory, provided one can arrange all anticommuting
ghosts in a relevant spinorial  representation space. One needs the equations
$\Phi_{\alpha}$ to contain gauge fields only (no scalars).  In this unifying
theory, all reminders to ghost number assigments disappear.
\lref\wati{W.~Taylor IV,
``D-brane field theory on compact spaces'',
hep-th/9611042, Phys.Lett. B394 (1997) 283-287}
Summarizing, the theory containing scalars like $\phi$ or $\varphi$ can be
pushed up in dimensions, making the scalars the remaining components of
the gauge field. This is completely parallel to the way
$T$-duality works for  the theories on $D$-branes \wati.

In the next section, we will apply these general statements to specific
theories, and discuss the  observables.

\subsec{Examples: Octonionic   theories}

\sssec{The \quad nine \quad dimensional \quad theory}
As an explicit example, let us look at the $9$-dimensional version
of what was done in the previous section, that is, the case $D=8$.
In what follows, for the sake of    notational clarity,    greek indices
$\mu, \nu,...$
run from $1$ to $9$, latin
indices
$i,j,...$
run from $1$ to $8$ and  greek
indices
$\alpha, \beta,...$
run from $1$ to $7$.

The $8$ dimensional theory  relies on the seven independent constraints
\eqn\octo{\Phi_{\alpha} = F_{8\alpha} -  c_{\alpha\beta\gamma} F^{\beta\gamma}}
where the $c_{\alpha\beta\gamma}$
are the structure coefficients of octonions \bks.
Its field content is that of $\CN=1$ $d=10$ theory dimensionally reduced
down to nine dimensions. The fermions are $\psi_{\mu}$ in $\bf 9$ of
$SO(9)$
and
$\chi_{\alpha}$ in $\bf 7$ of $Spin(7) \subset SO(8) \subset SO(9)$ \bks.

Assuming that space-time nine-fold splits: $M^{9} = {\bf S}^{1} \times
\CX^{8}$, with $\CX^{8}$ being $Spin(7)$ manifold\foot{The requirement that
$\CX^{8}$ is $Spin(7)$ manifold can be relaxed, see below},
the Lagrangian found by the standard procedure is:
\eqn\lg{\eqalign{& L =  \{ Q,
\int {\Tr} \biggl( i \chi^{\alpha}
(F_{8\alpha} -  c_{\alpha\beta\gamma} F^{\beta\gamma}  + {{ie^{2}}\over{2}}
H^{\alpha}) -
{1\over{e^{2}}} \psi_{\mu} ( F_{\mu 9} - i D_{\mu} \varphi ) \biggr) \} \cr}}
where  we omit the space-time
metric.
The expression of the action of $Q$ is defined in \nilgen.

This $Q$ invariant action can be expanded after standard
elimination
of $H$ as:
\eqn\lgg{L =   - {{1}\over{e^{2}}} \int {\Tr}( | F_{8\alpha}|^{2}
+|F_{\alpha\beta}|^{2} +| F_{\mu 9}|^{2} + |D_{\mu} \varphi|^{2}
+\ldots ) \biggr) }
that is
\eqn\lgrnn{L =  - {{1}\over{e^{2}}} \int {\Tr}(| F_{\mu \nu}|^2 + |D_{\mu} \varphi|^2
+\ldots  )}
where   $\ldots$ stand for $d$-exact terms, topological terms
and fermions.

The bosonic part of the action \lgg\ is invariant under $SO(9)$ rotations.
One easily finds that by the dimensional reduction
and  the  field redefinitions detailed in the previous section, that
\lgrnn\ can be dimensionally reduced to the  $8$
dimensional CohFT action built in \bks\
(for the $J$ case,
corresponding to $Spin (7)$ holonomy).

The link between  this $9$-dimensional theory and  the
$10$-dimensional super Yang-Mills theory  is quite transparent:
$|F_{\mu \nu}|^2 + |D_{\mu} \varphi|^2$
is the   dimensional reduction of the
$10$ dimensional Yang Mills action and  the number of fermions in the
$9$ dimensional theory, that is, $9$ components for the $\psi_\mu$ and
$7$
components for
the $\chi_\alpha$,
counts the $16$ independent
components of the Majorana-Weyl
spinor which is the
$\CN=1$ supersymmetric partner of the Yang-Mills field in $10$ dimensions.

Notice that the space of allowed eightfolds $\CX^{8}$ is not bounded
by Riemannian manifolds of $Spin(7)$ holonomy. One may replace the
equations
\octo\ by their deformations, which may involve other geometrical
structures. The link to supersymmetric Yang-Mills theory is more
involved in this case.

\sssec{Seven \quad dimensional \quad theory.}  In \bks\
an eight dimensional   CohFT is constructed using the gauge fixing of the
topological invariant
\eqn\sc {\int_{\CX^{8}} \Omega^{4} \wedge {\Tr} \  F \wedge F.}
 Therefore,
a meaningful  question is  whether a seven dimensional QFT exists,
which is directly defined
from the Chern-Simons action associated to previous action
\eqn\CSsss{\eqalign{
&    \int_{M^{7}} \Omega^{4} \wedge {\Tr} 
(A\wedge d A +{2\over{3}} A\wedge A\wedge A),
 \cr}}
that is,
\eqn\CSs{\eqalign{
&    \int_{M^{7}}
c_{\alpha\beta\gamma} \ {\Tr} (A_{\alpha}\p_{\beta}
 A_{\gamma}  +{2\over{3}} A_{\alpha} A_{\beta}  A_{\gamma} ).
 \cr}}
The quantization of this action is however unclear, in contrast
with the
 $3$-dimensional case,  
for  Gauss law in the $A_7=0$ gauge is 
not enough to consistently solve  the theory.

It is therefore more adequate to consider a simpler action, assuming that
$M^7\sim {\bf S}^{1} \times \CX^6$, with $\CX^{6}$ being Calabi-Yau threefold.
As expected, the theory will be related to the CohFT  relying on
balanced version of Donaldson-Uhlenbeck-Yau equations,
which are  dimensionally reduced octonionic instanton equations
 (see \bks). The effective six-dimensional theory turns out to be
a six-dimensional version of gauged WZW model.

We formulate the theory for $\CX^{6}$ K\"ahler threefold (not necessary
Calabi-Yau). The local coordinates on $\CX^{6}$ will be denoted as $z^{i}, \zb^{\ib}$.
The field content is the twisted version of the
field content of dimensionally reduced $9$-dimensional theory \lg. 
In other words the bosonic
fields are $(0,1)$ and $(1,0)$ forms $A_{i}, \bar A_{\ib}$ as gauge fields, $(3,0)$ and
$(0,3)$ forms $\phi, \bar\phi$
as Higgs fields in the adjoint representation and the
time-like component $A_{t}$ of the gauge field and its friend real scalar
$\varphi$. The fermions are: $(0,0)$ forms $\chi_{0}, \psi_{t \equiv 7}$, $(2,0)$ and
$(0,2)$ forms $\chi_{ij}dz^{i}\wedge dz^{j}$ and its conujugate, the 
$(3,0)$ and $(0,3)$ forms  $\psi_{\phi}$ and its conjugate
and $(1,0)$ and $(0,1)$ forms $\psi_{i} dz^{i}$ and its conjugate. 
This collection of fields
is suitable for posing the moduli problem \bks.

The six-dimensional equations are:
\eqn\duy{
F^{0,2} = \pb_{A}^{\dagger} \bar \phi, \quad \omega^{2} \wedge
F^{1,1} \equiv (g^{i\bar i} F_{i\bar i} )\omega^{3} =  [\phi, \bar\phi], }
where $\omega = \omega_{i\bar j}dz^{i} \wedge d{\bar z}^{\bar j}$
is the K\"ahler form on a six-fold $\CX^{6}$.
The Lagrangian found by the procedure described above
is:
\eqn\lgrn{\eqalign{
& L =  \{ Q \int {\Tr} \biggl( 
\chi_{0}( \omega^{2} \wedge F +  [\phi,
\bar\phi] + {{ie^{2}}\over{2}} H_{0})\cr & + \chi_{\bar i \bar j}(
{{ie^{2}}\over{2}} H_{ij} - F_{ij} - \epsilon_{ijk}D_{\bar k}\phi) + c.c. +
\cr & + \left( \psi_{\phi} [ D_{t} - i\varphi , \bar\phi] + 
\bar\psi_{\bar \phi} [ D_{t} - i\varphi, \phi] \right) + \psi^{\mu} ( F_{t\mu} + i D_{\mu} \varphi )  \biggr) \}, \cr}}
where
$\mu = 1, \ldots, 7$, $i,j,k = 1,2,3$, and we omit the metric $g_{i\bar k}$ in
all our formulae.
The action of $Q$ on $\phi, \bar\phi$ is obtained by dimensional reduction:
\eqn\qfi{\matrix{Q \phi = \psi_{\phi} & 
Q \psi_{\phi} = D_{t} \phi + i [\varphi, \phi]\cr 
Q \bar\phi = \bar\psi_{\bar\phi} & Q \bar\psi_{\bar\phi} = 
D_{t} \bar\phi - i [\varphi, \bar\phi]\cr}}
After expansion of $\{ Q, \ldots \}$ the action \lgrn\ is
similar in form to \lgrnn.

\newsec{Observables}

\subsec{Observables in $\CH_{D}$ theory}
One may construct the observables in the theory
by means of the descend procedure: start with the operator
$\CO^{0}_{k} = {\Tr} \phi^{k}$ and compute its exterior derivative
$d\CO^{0}$. It is $Q$ of a one-form valued operator called $\CS^{1}_{k}$.
The integral
of $\CS^{1}_{k}$ along a closed curve $C$ is therefore a
$Q$-closed observable $\CO^{1}_{k}$. Take $d$ of $\CS^{1}_{k}$ and so on:
$$
d \CS^{l}_{k} = \{ Q , \CS^{l+1}_{k} \}
$$
One gets in this way a chain of observables $\CO^{p}_{k}$:
$$
\CO^{p}_{k} = \int_{C^{p}} \CS^{p}_{k}
$$
where $C^{p}$ is a $p$-cycle  and a map
\eqn\dnlsmp{\mu_{k}: H_{*}(\CX^{D}; \IR) \to H^{2k - *} ({\CM}; {\IR}) }
which in the field theoretic language is our desired map
$\CO_{i} \mapsto \omega_{i}$.
The form $\omega_{i}$ represnts a cohomology class of the
moduli space $\CM = \amalg_{n} \CM_{n}$. In this paper
we   ignore the subtleties
associated  to
intersections of cycles $C^{l}$, over which $\CS^{l}$ are integrated.

\subsec{Observables in $\CK_{D}$ theory}

On can ask whether there are in $D+1$ dimensions operators
which represent non-trivial classes of  $Q$-cohomology.
The answer to this question is positive. This is a  generalization
of    the $3$ and $5$ dimensional cases studied in \nikitathes.
Recall the
property \nil\ that
the supercharge $Q$ squares to the combination of the $t$ translation
and the gauge transformation with parameter $A_{t} + i\varphi$.
There are two types of observables in the theory.

The first type
observables are obtained by the standard descend procedure applied to the
gauge invariant zero-observables:
\eqn\zrobsr{{\CO}^{0}_{k} = {\Tr}g^{k}, \quad g = P\exp\oint (A_{t} +
i\varphi)dt}
which are the analogues of ``vertical'' Wilson loops.

The observables of second type contain Chern-Simons terms:
start with
\eqn\CSi{L_{CS} =  \int_{M^{D+1}}  T^{D-2} \wedge {\Tr} (A\wedge d A +{2\over{3}}
A\wedge A\wedge A)} where
$T^{D-2}$ is a closed $D-2$-form coming from $\CX^{D}$, i.e. it is a pullback
of a form on $\CX^{D}$ with respect to the projection
$M^{D+1} \to \CX^{D}$.
The following operator turns out to be $Q$ invariant:
\eqn\CSiii{\CL_{CS,1} = \int_{M^{D+1}} T^{D-2} \wedge \left( CS_{3}(A) +
2 {\Tr} (i\varphi F + \psi \psi ) \wedge dt \right)}
This operator is actually a cousin of the term \alp\ in the SQM.
As opposed to the $Q$ exact Lagrangian  \lg\
which is the analogue of \rgltr, $\CL_{CS} $ is not  $SO(D+1)$
 invariant. Rather  it is only
 $SO(D)$ invariant, as it involves a form $T^{D-2}$ on $\CX^{D}$.

More generally, given a closed $D-2p$ - form on $\CX^{D}$ one may construct
the observable of the following type:
\eqn\CSiv{\CL_{CS, p} = \int_{M^{D+1}} T^{D-2p} \wedge \left( CS_{2p+1}(A) +
(p+1) {\Tr} (i\varphi F^{p}   + \sum_{l=0}^{p-1}  \psi F^{l} \psi F^{p-1-l})
\wedge dt \right)}
with $CS_{2p+1}(A)$ being the standard Chern-Simons $2p+1$-form:
\eqn\CSv{\eqalign{& CS_{3}(A) = {\Tr} \left( AdA + {2\over{3}} A^{3}
\right)\cr
& CS_{5}(A) = {\Tr} \left( A(dA)^{2} + {3\over{2}} A^{3}dA  +
{3\over{5}} A^{5} \right)\cr
& CS_{2p+1}(A)  = (p+1) \int_{0}^{1} s^{p}ds {\Tr} \left(
A ( dA + s A^{2})^{p} \right)\cr} }

All these operators share the
property  of being only $SO(D)$ invariant.

\subsec{Quantization of $T^{D-2p}$.}
The forms $T^{D-2p}$ are the higher
dimensional analogues of the level ${\bf k}$ in the three dimensional
Chern-Simons theory. One expects an analogue
of the quantization condition of ${\bf k}$, like the quantization of Kahler
form $\omega$ in \avatar. Indeed, the presence of
the term $T^{D-2p} \wedge CS_{2p+1}(A)$ implies that in order to
preserve gauge invariance the forms $T^{D-2p}$ must represent integral
cohomology classes of $\CX^{D}$:
\eqn\quan{[ T^{D-2p} ] \in H^{D-2p} (\CX^{D}; (2\pi i)^{p+2} \IZ)}
and the operators \CSiv, \CSv\ make sense only
when they are in the exponential.

So, the actual observables of second type are:
\eqn\actobs{\CO_{p}(T^{D-2p}) = \exp ( \CL_{CS,p} ).}

\subsec{Flow to
``genuine'' Chern-Simons theory}

In $D+1$ dimension we systematically consider an action which is
the sum of
a Chern-Simons like action as in \CSiii\  and a $Q$ exact action as in \lg.
This leads to a well defined QFT.  However one may wonder about  the relation
of this theory to a genuine  Chern-Simons theory, without  supersymmetric
terms.

\sssec{Getting \quad rid \quad of \quad }$\varphi$.

It is possible to map the observable \actobs\ for $p=1$ to more
conventional Chern-Simons like action. The idea
is the following.
\lref\Wittrev{E.~Witten, ``Two dimensional gauge theory revisited'',
hep-th/9204083, J. Geom. Phys. 9 (1992) 303-368 }
Suppose that the representative of $T^{D-2}$ is such
that
\eqn\constrcs{T^{D-2} \wedge F \wedge dt = 0} is actually
one of the constraints, say $\Phi_{r}$. Then consider adding to the
Lagrangian \newlagran\ the term
\eqn\trick{\Delta \CL = \kappa \{ Q, \int  {\Tr}\  \left(
\chi_{r} \varphi \right) {\rm vol}_{g} \} }
where $\kappa$ is an arbitrary  coefficient. Since the constraint
$\Phi_{r}$
is indirectly imposed by varying $A_{t}$ it may seem that taking
the limit $\kappa \to \infty$ does not change the
behavior of the theory. The advantage is that in the limit $\kappa =
\infty$
we can forget about other terms in \newlagran\ where $\varphi$ appears
and integrate out $\varphi, \psi_{t}, \chi_{r}$ and $H_{r}$ altogether.
This argument is very similar to the explanation
of the relation between the physical and topological Yang-Mills theories
in two dimensions, proposed in \Wittrev.
In this way we end up with the Chern-Simons theory together
with second class constraints $\Phi_{\alpha} = 0 , \alpha \neq r$.
In fact, the constraints are imposed in the $e^{2} \to 0$ limit.

For example, in dimensions $D=2,4,8$ the equations
can be written as follows:
$\Phi_{\alpha} = F_{D\alpha} - \xi_{\alpha\beta\gamma}
F^{\beta\gamma}$ where:
\eqn\str{\eqalign{D = 2 \quad & \quad \xi_{\alpha\beta\gamma} = 0\cr
D = 4 \quad & \quad \xi_{\alpha\beta\gamma} = \epsilon_{\alpha\beta\gamma}\cr
D = 8 \quad & \quad \xi_{\alpha\beta\gamma} = c_{\alpha\beta\gamma}\cr}}
In this cases $r=D-1$.

\sssec{Strong \quad coupling \quad limit.}
Now let us formally study the
opposite,
strong coupling limit $e^{2} \to \infty$. The following manipulations
are not justified without proper regularization. They may serve as
useful indications but not as a proof that one may be
left with the Chern-Simons theory alone.
Indeed, if the term $\{ Q, \chi^{\alpha} \Phi_{\alpha} \}$
in \newlagran\
can be neglected then we end up with the action (provided that $\Phi_{r} =
T^{D-2} \wedge F \wedge dt$):
\eqn\axial{\eqalign{
&\int      T^{D-2} \wedge dt \wedge {\Tr}
\left(
A \wedge \p_{t}  A   +(A_{t}+i\varphi) \wedge F +{\half} \psi \wedge \psi
\right)
 \cr
+& \{ {\lies}, i \int  {d^{D+1}x}  \  {\Tr} \left( \chi_r \cdot \varphi
+ {{ie^{2}}\over{2}} \sum_{\alpha\neq r} \chi^{\alpha}  H^{\alpha} \right) \} \cr
  +& \{ {\lies}, \int  d^{D+1} x  \  {\Tr} (\bar c  A_{t}^{\perp})\}
 \cr}}
where we  introduce the multiplet $(\bar c , \lambda)$
of Faddeev-Popov anti-ghost and Lagrange multiplier
for the gauge $A_{t}^{\perp} = 0$, where $A_{t}^{\perp}$ is the projection
of $A_{t}$ onto the complement to the Cartan subalgebra plus
the projection of the abelian part of $A_{t}$ onto the space of
non-zero (in $t$) modes.
The supercharge $\lies$ acts on them as follows:
$$
{\lies} \bar c = \lambda - [c, \bar c], \quad Q \lambda = D_{t} \bar c +
i[\varphi, \bar c] - [ c, \lambda]
$$
Expanding \axial\ one formally eliminates all the fields except
$\chi_{\alpha}, \alpha \neq r$, $A_{i}, i = 1, \ldots , D$, $\bar c$
and $c$ which have the action:
\eqn\axiall{
\int dt \wedge \left( T^{D-2}  \wedge
{\Tr} ( A \wedge  D_{t} A ) + {\Tr} (\chi^{\alpha} \star D_{t}
\chi^{\alpha}) + {\Tr} {\bar c} \star D_{t} c \right) }
with $D_{t} = \p_{t} + [ A_{t}^{(0)}, \cdot ]$, $A^{(0)}_{t}$ being the
constant in $t$ Cartan-valued matrix. The Hodge star $\star$ is taken in
$D$ dimensions. Formally the determinants cancel
in topologically trivial  backgrounds.
Of course, non-trivial gauge backgrounds induce effective action and
then it may be not possible to neglect the ${1\over{e^{2}}} F_{\mu\nu}^{2}$
terms.

\subsec{Discussion of breaking the $SO(D+1)$ Lorentz invariance}

One may complain about the breaking of Lorentz invariance by
the operators \CSiv. In fact, as long as $2p < D$, the operator
\CSiv\ is an observable, moreover the form
$T^{D-2p}$ can be taken supported at some submanifold $\Sigma^{2p}$.
As such, it is not illegal for the observable to violate
$D+1$ dimensional Lorentz invariance, since the choice
of $\Sigma^{2p}$ already breaks it. What is a little bit
surprising is that the choice of supercharge $\lies$ or $Q$ which preserves
the observable violates $2p+1$ dimensional Lorentz invariance,
as it requires the choice of $t$ direction.
\lref\seibfi{N.~Seiberg, ``Five Dimensional SUSY Field Theories, Non-trivial Fixed Points and String
Dynamics'', Phys.Lett. B388 (1996) 753-76, hep-th/9608111}
On the other hand, if $2p = D = 4$ then the Chern-Simons
part of the operator \CSiv\ is being integrated over whole
$4+1$ -dimensional space-time. It is possible
to check that when the $Q$-exact regulators are
included added we end up with Lorentz invariant
action in $\IR^{4+1}$  corresponding to five dimensional
super-Yang-Mills with running coupling ${1\over{g^{2}}} \sim \varphi$
discussed in \seibfi.
Note
that a proposal that infrared fixed points in five dimensional
supersymmetric gauge theories are described by Chern-Simons theories
has been made in  \kvd.

In fact, all this puts the constructions of \shatashi\whatwethink\
discussing five-dimensional Chern-Simons field theories in a more
solid context.

\newsec{More examples}
 Since pure topological  Yang Mills theories exist in $4$ and $2$
dimensions,
we can study the associated five and three  dimensional theories.

\subsec{Five dimensional theory}

This theory was studied in \avatar\nikitathes.
It has the field content
of the partially twisted $\CN=1$ $d=6$ theory, dimensionally reduced
down to five dimensions.

\subsec{Three dimensional theory and Verlinde formula}

The interesting property of this theory is that it has the field
content of the partially twisted $\CN=1$ $d=4$ theory, dimensionally
reduced down to three dimensions. The expectation value of the
observable containing the three dimensional Chern-Simons action is the
Verlinde formula for the number of conformal blocks in
two dimensional WZW theory \nikitathes. In this case the trick \trick\
with eliminating $\varphi$ allows one to get rid of all fields
except $A$ and establishes the equivalence
to the ordinary three dimensional Chern-Simons theory.

\subsec{Theories with matter}

\lref\mverl{R.~Dijkgraaf, E.~Verlinde, H.~Verlinde, ``Matrix String
Theory'',
hep-th/9703030}
\lref\mseib{O.~Aharony, M.~Berkooz, S.~Kachru, N.~Seiberg and
E.~Silverstein,
``Matrix description of interacting theories in six dimensions'',
hep-th/9707079}
\lref\mwitt{E.~Witten, ``On The Conformal Theory of The Higgs Branch'',
hep-th/9707093}
\lref\mlns{A.~Losev, S.~Shatashvili, N.~Nekrasov, in progress}
The theories in five and three dimensions which
have the field content of $\CN=1$ $d=6$ and/or $d=4$ super-Yang-Mills
can be coupled to matter multiplets. In this way one gets
a deformation of Higgs branches of  theories by including
one extra dimension compactified on a circle. The simplest example is
the theory in $D=2$ describing gauged linear sigma model. Its Higgs
branch (more precisely, its effective low energy
target space) $\CM_{H}$ must be a K\"ahler manifold due to  supersymmetry.
By using our trick of going one dimension higher we obtained a
deformed theory, where certain  correlation
functions contain insertions of expansion $\hat A(\CM_{H})$.
Notice that the
first non-trivial operator in the expansion of $A$-genus
has dimension four. It implies that effectively the target space
$\CM_{H}$ is changed in codimension four. It is very tempting to
speculate that this codimension four change is related to
recent proposals of incorporating stringy interactions in
Matrix theory \mverl\mseib\mwitt.

\newsec{Application: new theory in $D$ dimensions}

Once we have constructed a theory in $D+1$ dimensions with
supercharge, squaring to the translation in the $D+1$'st direction
we may consider its compactifications.

The dimensional reduction gives us back the theory in $D$ dimensions
we started from, but compactification on a circle produces a
deformation of the original theory, the radius $R$ being the
parameter of the deformation. As a guiding example let us start
in $D=2$.  The theory describing moduli space of flat connections is
$2d$ topological Yang-Mills theory. The corresponding theory in $2+1$
dimensions is going to be twisted $3d$ SYM with the Chern-Simons
action as an observable. The compactification on a circle
of the latter produces gauged $G/G$ WZW theory \btverlinde\gerasimov,
which coincides
with $2d$ YM in the large $k$ limit. The compactification
on an interval gives rise to the WZW model itself. See \fadsha\ for
the similar construction in the bosonic $D=4$ case and 
\elitzur\rf\baugros\bau\wittjones\vrlsq\
for bosonic $D=2$ case.

The same procedure works in higher dimensions. 
One has to be careful,
as the compactification works nicely only for the modes, annihilated
by the supercharge $\lies$. In particular, all constraints $\Phi_{\alpha}$
but one must be imposed. In addition, one has to solve the equation
$$
F_{\mu t} + iD_{\mu} \varphi = 0
$$
which states that
\eqn\bound{A \vert_{t=\tau} = \left( A \vert_{t=0} \right)^{g(\tau)} }
where
\eqn\pexp{g(\tau) = P\exp\int_{0}^{\tau} (A_{t} + i\varphi)dt}

The $CS_{3}$ type observable gives rise to the $WZW_{2}$ like action.
The observables corresponding to $CS_{5}$ and so on produce
the generalized $WZW$-like actions, computed in \clash.

One can also see these relations in the canonical approach. Let us consider
for simplicity the case $D=8$ with equations $\Phi_{\alpha}=0$ being the
complexified
instanton equations (case $H$ in the terminology of \bks).
Then $\CX$ is K\"ahler manifold.
Let $T^{6} = \omega^{3}$,
$\omega$ being the K\"ahler form.
In the  weak coupling limit (after eliminating $\varphi, \chi_{1}$ and
$H_{1}, \psi_{t}$) the effective space $\CB$ of fields
is the space of solutions
to equations $\Phi_{\alpha} = 0 , \alpha > 1$. Suppose that
we have inserted an operator $\CO_{1} (T^{6})$. The form $T^{6}$
defines a symplectic form on the space of gauge fields on $\CX^{D}$:
\eqn\sympld{\Omega = \int_{\CX^{8}} T^{6} \wedge {\Tr} \delta A \wedge
\delta A}
which is invariant under the gauge group action. Now we may
attempt to quantize the symplectic quotient of $\CB$ by the action
of gauge group. The Gauss law:
$$
T^{6} \wedge F = 0
$$
translates to the following property of the wavefunctional
(we work in holomorphic polarization):
\eqn\prop{
\Psi ({\bar A}^{g}) = \exp \left( \int_{{\CX}^{8}} T^{6} \wedge \left(
S_{WZW_{2}}(g) + {\Tr} ( {\bar A} g^{-1} \p g)  \right) \right) \Psi ( \bar A)
}
where $\bar A$ is the $(0,1)$ part of the gauge field. It must belong
to $\CB$, i.e. to obey $(0,2)$ part of the complexified instanton equations.
One may find a formal solution to \prop\ as a path integral
in $WZW_{8}$ theory in the background gauge field $\bar A$
(this is completely parallel to \gerasimov\avatar\clash):
\eqn\frmlsl{\Psi ({\bar A}) \sim
\int Dg \exp\left( \int_{{\CX}^{8}} T^{6} \wedge \left(
S_{WZW_{2}}(g) + {\Tr} ( {\bar A} g^{-1} \p g)  \right) \right)  }

The condition $\bar A \in \CB$ which must be imposed on the
wave function by hand looks a little bit unnatural. It is possible
to represent this condition in a different way, which makes the
interesting use of the supersymmetric quantum 
mechanics which we described
earlier.

Consider $D=6$ case with the equations \duy. Let us again separate
them as $\Phi^{0,2} = 
F^{0,2} - \pb^{\dagger}_{A} \bar\phi $ and
``Gauss law'' $\mu = \omega^{2} \wedge F^{1,1} - [ \phi, \bar\phi] $.
Consider the canonical quantization of the theory with the Lagrangian
\lgrn.
The bosonic fields in the theory form a configuration space $\CA$ which
is the space of gauge fields $A$ times the space of $(3,0)$
$\lieg_{\IC}$-valued
forms $\phi$.
The space $\CA$ is an infinite-dimensional K\"ahler manifold,
the K\"ahler form being
\eqn\klrfrm{\Omega = \int_{\CX^{6}} \omega^{2} \wedge 
{\Tr} \delta A \wedge \delta \bar A   + {\Tr} \delta \phi 
\wedge \delta \bar\phi}
It is preserved by the gauge group action and $\mu$ is the corresponding
moment map. The fermionic kinetic term is roughly
\eqn\frmknt{{\Tr} \left( 
\psi_{i} D_{t} \psi_{\ib} + \chi_{ij} D_{t} \chi_{\ib\jb} + 
\eta D_{t} \eta + \chi_{0} D_{t} \chi_{0} \right)}
where the metric is implicit. 
 This kinetic term suggests that the wave functional can be factorized
as follows:
\eqn\fctrz{
\Psi = \Psi_{0} (A, \bar A, \phi, \bar\phi, \chi_{ij} , \psi_{\ib}) \otimes  v }
where $v$ is a vector in the two dimensional Hilbert space obtained by quantizing $\chi_{0}, \eta$ system.
The piece $\Psi_{0}$ is naturally a section of
$ \Lambda^{0,*} ({\CA}) \otimes \Lambda^{*}E $ 
where $E$ is the complex
vector 
bundle over $\CA$ whose fiber over $(A, \phi)$ is the space $\Omega^{0,2}_{\lieg}$.
The supercharge $\lies$ can be represented as  the equivariant version of
the infinite-dimensional operator
\eqn\opr{
\pb + \pb^{\dagger} + \delta + \delta^{\dagger}
}
where $\delta$ is the Koszul operator (see, \GrHa) , which maps 
$\Gamma( \Lambda^{p}E) \to \Gamma( \Lambda^{p+1}E)$ by exterior
multiplication by $\Phi^{0,2} \in \Gamma (E)$. 

The kernel of $\lies$ is identified with the equivariant
cohomology of the operator $\pb  + \delta + \varphi \iota_{V}$,
where $V$ represents the complexified gauge group action. 
By the standard spectral sequence
techniques one may first compute the cohomology of $\delta$ which is
roughly speaking equivalent to imposing the $\Phi^{0,2} =0 $ constraint.
Due to the infinite-dimensionality of the problem it
seems more adequate to work with Koszul operator rather then
with imposed non-linear equations $\Phi^{0,2}=0$. 
The last remark concerns the role of $\mu$. It is known that taking the
quotient with respect to the complexified group in the sense of theory 
of invariants is equivalent to imposing $\mu =0$ constraint first and
then taking the quotient with respect to the compact group.
\lref\kirwan{F.~Kirwan, ``Cohomology of quotients in symplectic
and algebraic geometry'', Math. Notes, Princeton University Press, 1985\semi
L.C. Jeffrey, F.C. Kirwan,
``Localization for nonabelian group actions'', alg-geom/9307001}
Cohomology-wise there is a surjective map from the equivariant
cohomology of $\CA$ to the cohomology of the symplectic
quotient $\CA //\CG$ \kirwan. 
This explains why we need not include
$\mu$ in the Koszul complex \Witdgt. 

The analogous construction can be presented in other cases considered in 
this paper as well.

\newsec{Discussion and conclusions}

We have constructed twisted supersymmetric theories in higher dimensions.
In fact, we only described one relevant supercharge $Q$. It is clear,
though, that one may write down the superalgebra involving all
supercharges. Of course, on a curved background ${\bf S}^{1} \times
{\CX}^{D}$
only one supercharge is conserved.

The interesting property of this supersymmetry is that it allows
to construct $Q$-invariant observables of Chern-Simons type. Pure
Chern-Simons theory is not very well defined and needs a regularization.
The supersymmetric gauge theory provides such a regularization.
The case of three dimensional theory is exceptional in the sense that
one can get rid of all the regularizing fields by going to
 well justified strong coupling limit (see \bau\ for the formal derivation of the coupled
 Chern-Simons Yang-Mills system in three dimensions). This is not so in higher
dimensional cases where the extra constraints are important.

Three dimensional Chern-Simons theory induces two dimensional
WZW model. It turns out that a similar statement holds in higher
dimensional case provided that one works in $Q$-cohomology. This
may provide a further justification of higher dimensional
WZW theories \avatar\clash.

\lref\GHM{M.~Green, J.~Harvey, G.~Moore,
``I-Brane Inflow and Anomalous Couplings on D-Branes'', hep-th/9605033,
Class.Quant.Grav. 14 (1997) 47-52}

The theories which we have studied in our paper may be of some
relevance in the context of theories on $D$-branes. Indeed,
Chern-Simons-like couplings to $RR$ fields are known to be present
in the effective actions on the worldvolumes of $D$-branes \GHM.
Also, Chern-Simons terms appear on the worldvolumes of euclidean
\lref\molough{M.O'Loughlin, ``Chern-Simons from Dirichlet $2$-brane
instantons'', hep-th/9601179,
Phys.Lett. B385 (1996) 103-108}
$D2$ branes, wrapping supersymmetric three-cycles \molough.

Finally, the process of going one  dimension up is similar to
the action of $T$-duality on the worldvolume theory of a transverse
$D$-brane.

As our paper was ready for publication we have learned about the
recent preprint \afsl\ which also drew attention to seven dimensional
theory with Chern-Simons action.

\centerline{\bf Acknowledgements}

We are grateful to M.~Douglas, A.~Gerasimov, E.~Martinec, G.~Moore,
A.~Rosly, S.~Shatashvili and C.~Vafa for interesting discussions
on the subjects related to this work.

The research of A.~L.~ is supported partially by DOE under grant
DE-FG02-92ER40704, by
PYI grant PHY-9058501 and RFFI under grant 96-01-01101.
The research of N.~N.~ is supported by Harvard Society of Fellows,
partially by NSF under  grant
PHY-92-18167,  and partially by RFFI grant 96-02-18046.
A.~L.~ and N.~N.~ are also supported by
grant 96-15-96455 for scientific schools.

\listrefs
\bye